\documentclass[floatfix,reprint,amsmath,amssymb,aps,prl]{revtex4-1}
\usepackage{times,mathptmx}
\usepackage{graphicx}
\usepackage[english]{babel}
\usepackage{epstopdf}
\usepackage{xcolor}

\begin{document}

\title{Kink-based mirrorless quasi-bistability in resonantly absorbing media}

\author{Denis~V.~Novitsky$^{1,2}$}
\email{dvnovitsky@gmail.com}
\author{Alexander~S.~Shalin$^{2,3}$}

\affiliation{$^1$B.~I.~Stepanov Institute of Physics, National
Academy of Sciences of Belarus, Nezavisimosti Avenue 68, 220072
Minsk, Belarus \\ $^2$ITMO University, Kronverksky Prospekt 49,
197101 St. Petersburg, Russia \\ $^3$Ulyanovsk State University, Lev Tolstoy Street 42, 432017 Ulyanovsk, Russia}

\date{\today}

\begin{abstract}
Optical bistability, the basic nonlinear phenomenon mediating the control of light by light, paves the way to the all-optical logic being of ultimate demand for a plethora of applications in laser information technologies. The desirable features of the optically bistable elements are low power consumption, speed of switching and small size. The two most general designs are driven by the presence or absence of an external feedback giving rise to a variety of possible setups. Among them, mirrorless architecture seems promising being free of bulky mirrors, resonant cavities, photonic crystals, etc. In this paper, we propose a novel method to achieve optical quasi-bistability governed by the formation of specific nonlinear waveforms called “kinks”. We show that a thin layer of the relatively dilute resonant medium specially designed to support kinks could serve as a platform for compact, ultra-fast, low power optical switching. This new physical mechanism do not require high densities of resonant particles specific for other feedback-free devices driven by local field corrections and dipole-dipole interactions, and enhance the overall practical relevance of such devices for optical computing.
\end{abstract}

\maketitle

Bistability lies at the heart of the digital information storage and processing systems. This justifies the importance of the search for the bistability in different optical systems considered as promising elements of future optical computers \cite{Karim,Li}. Bistable response of an optical system implies the possibility of the two stable states at the same intensity and polarization of the incident radiation. These states differ in the output field intensity or polarization. Leaving aside the polarization bistability, the main feature of the bistable response is the hysteresis loop in the dependence of the output intensity on the input one.

The standard approach to optical bistability was based on nonlinear media with a feedback \cite{Gibbs}. The feedback can be realized in a number of ways. The earliest and the most straightforward approach is to use the mirrors obtaining a nonlinear-medium-filled Fabry-Perot resonator \cite{Gibbs1976,Felber1976}. The distributed feedback was proposed later to realize bistability as well \cite{Winful1979}. In recent years, more advanced realizations of the feedback were studied, such as Bragg reflectors \cite{Acklin1993}, photonic crystals \cite{Lidorikis2000,Soljacic2002,Teimourpour2012,Sreekanth2015}, photonic crystal waveguide/cavity systems \cite{Yanik2003,YanikOL2003,Mingaleev2006,Kabakova2010,Zhang2010}, and metamaterials \cite{Lapine2014,Tuz2010,Tuz2012,Kim2018}. Moreover, bistability can be observed even in disordered media due to Anderson-localization-induced cavities formation \cite{Shadrivov2010,Yuan2016}. Although the cavity of a photonic-crystal bistable system can be even less than the light wavelength \cite{Soljacic2003}, the structure itself should contain several periods and is not so compact.

Fundamentally different mechanisms are utilized in the so-called mirrorless (or cavity-free) optical bistability. In this case, bistable response is generated only by the medium's nonlinearity, without need of any external feedback. In other words, the medium is bistable locally, at every given point, as in the case of magnetic hysteresis. The mirrorless bistablility can be induced, for example, by nonlinear absorption in semiconductors \cite{Rozanov1981,Miller1984,Gribkovskii1988}. Moreover, there is another interesting mechanism of mirrorless bistability, when resonantly absorbing particles (atoms, molecules, quantum dots) are packed closely enough to have many of them in a qubic wavelength. In such a dense resonant medium, the particles start to interact strongly with each other. The dipole-dipole interactions between the particles can be described in terms of local-field correction leading to the Lorentz shift of the resonant frequency. The Lorentz term introduces the additional nonlinearity into the optical Bloch equations, which are used to describe the medium, and results in the local (or intrinsic) bistable response. This mechanism of mirrorless bistability was proposed in the pioneering works by Bowden et al. \cite{Hopf1984,Ben-Aryeh1986,Haus1988}; many details and modifications were later studied by other authors \cite{Friedberg1989,Benedict1991,Malyshev1997,Afanas'ev1998,Afanas'ev1999,Brunel1999,Novitsky2008}. This mechanism was applied to study the bistability in the collections of rare-earth ions \cite{Hehlen1994}, quantum dots \cite{Paspalakis2006}, and molecular aggregates \cite{Malyshev1996}.

The propagation of optical pulses through bistable systems, as a rule, results in nonlinear transformations of the form and duration of the pulses \cite{Gibbs,Kabakova2010}. In particular, the spatial discontinuities called kinks were reported as a side manifestation of the inhomogeneous distribution of the excitation and the local nature of the mirrorless optical bistability \cite{Gibbs1985,Lindberg1986}. In this paper, we turn this observation inside out and propose to use kinks in a resonant medium as a basis for bistable-like response. The kinks, or self-similar shockwaves, were predicted to exist in resonant media in the regime of incoherent light-matter interaction in Ref. \cite{Ponomarenko2010} and later numerically studied in one of our previous papers \cite{Novitsky2017}. Hereinafter, we show that these kinks can deliver mirrorless quasi-bistability with fast (sub-nanosecond) and low-intensity switching in a quite compact slab (about 2 wavelengths or less) of a diluted resonant media. This particular design not requiring cavities, resonators, dense doping with resonant atoms or particles, rather large material slabs or intensities, etc., could be extremely useful for a variety of optical applications being simple for fabrication and processing.

Let us start with a brief discussion of the standard mirrorless bistability in the dense resonant media due to local-field correction. This correction takes into account the dipole-dipole interactions between atoms and results in the so-called Lorentz redshift of the resonant frequency. In the two-level approximation, the optical Bloch equations taking into account this effect can be written as \cite{Bowden1993,Afanas'ev1998,Crenshaw2008}
\begin{eqnarray}
\dot{\rho}&=& i l \Omega w + i \rho (\delta + \omega_L l w) - \rho/T_2, \label{rho} \\
\dot{w}&=&2 i (l^* \Omega^* \rho - \rho^* l \Omega) -
(w-1)/T_1,
\label{w}
\end{eqnarray}
where $w$ is the population difference between the ground and excited states, $\rho$ is the microscopic polarization (atomic dipole moment), $\Omega=\mu E/\hbar$ is the Rabi frequency, $E$ is the electric-field amplitude, $\mu$ is the dipole moment of the quantum transition, $\delta=\omega-\omega_0$ is the detuning between the light carrier frequency $\omega$ and the frequency of the quantum transition $\omega_0$, $T_{1}$ and $T_{2}$ are the relaxation times of population and polarization respectively, $\omega_L = 4 \pi \mu^2 C/3 \hbar$ is the Lorentz frequency, $C$ is
the concentration of two-level atoms, $l=(n_d^2+2)/3$ is the local-field
enhancement factor, $n_d$ is the refractive index of the host medium, and $\hbar$ is the reduced Planck constant. The nonlinear term $\sim \rho w$ in Eq. (\ref{rho}) is due to dipole-dipole interactions and is the source of the bistable response of the medium. Indeed, one can easily see this in the stationary approximation ($\dot{\rho}=\dot{w}=0$) obtaining the cubic equation for the population difference,
\begin{eqnarray}
(1-w) [1+(\delta T_2 + \omega_L T_2 l w)^2] = 4 T_1 T_2 l^2 |\Omega|^2 w. \label{cubic}
\end{eqnarray}
This equation can have one or three real roots depending on the light intensity given by $|\Omega|^2$. In the range of intensities, where there are three roots, the system can be in one or another of stable state depending on the previous history of its dynamics. This type of mirrorless bistability has a threshold and appears only for large enough nonlinear terms in Eq. (\ref{cubic}). In the literature, one can find the specific values for this threshold, namely $\omega_L T_2 > 4$ (for $l=1$) \cite{Friedberg1989}. One can also find the analytic expression for the bistability existence threshold in the $(\delta T_2,\omega_L T_2)$ coordinates \cite{Novitsky2008}. This threshold poses a problem for practical realization of such mirrorless bistability requiring the resonant medium to be dense enough. Indeed, even for the two-level particles with relatively large dipole moments $\mu \sim 10$ D and short relaxation time $T_2 \sim 1$ ps (e.g., quantum dots), the concentration should be $C \gtrsim 10^{19}$ cm$^{-3}$, which is not always possible to provide. Note also that $T_2$ governs the steady-state establishment and should be as short as possible to obtain ultrafast switching, whereas $T_1 \gg T_2$ sets the time after which the system returns to the initial state and can be used again.

\begin{figure}[t!]
\includegraphics[scale=1., clip=]{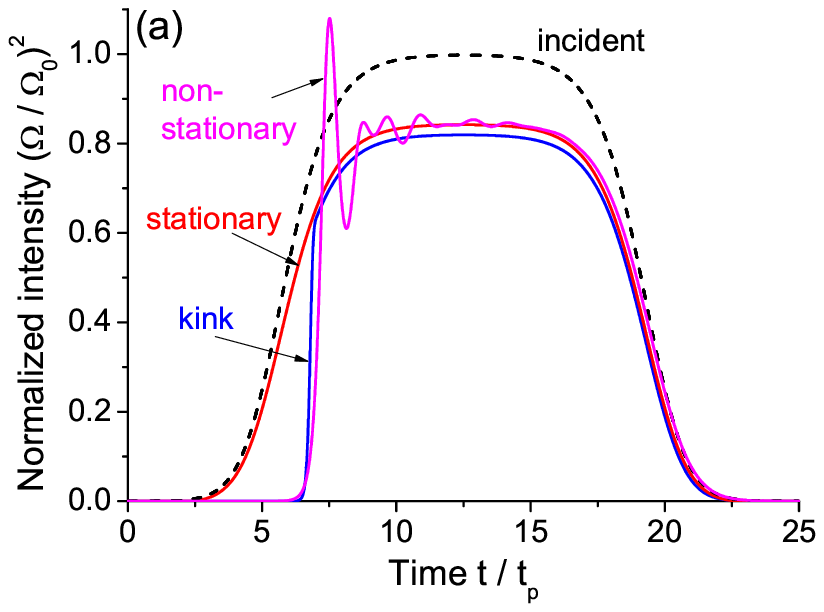}
\includegraphics[scale=1., clip=]{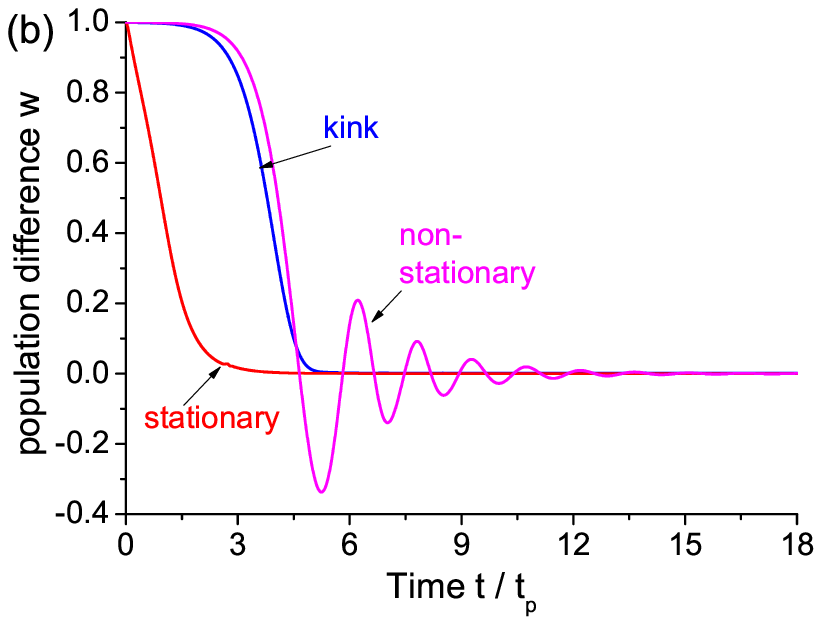}
\caption{\label{fig1} (a) Profiles of transmitted light intensity for the three cases discussed in the main text. (b) Corresponding dynamics of population difference at the entrance of the medium.}
\end{figure}

Here we propose another mechanism of mirrorless bistability based on kink formation in the resonant medium. This mechanism allows to alleviate the problems discussed above and obtain fast and low-intensity quasi-bistability based on the light interaction with the resonant two-level medium. We consider the case of the so-called kink-like pulses which form from the incident adiabatically switching waveform \cite{Novitsky2017},
\begin{eqnarray}
\Omega(t)= \frac{\Omega_0}{(1+e^{-(t-t_0)/t_p})(1+e^{(t-t'_0)/t_p})}, \label{waveform}
\end{eqnarray}
where $\Omega_0$ is the amplitude of the cw field (plateau), $t_p$ is the switching time, $t_0=5t_p$ and $t'_0=4 t_p$ are the offset times which govern the moments of field switch-on and switch-off, respectively. The profile given by Eq. \ref{waveform} is shown in Fig. \ref{fig1} with the dashed curve. The key condition for kink formation is $T_2 \ll t_p \ll T_1$, which evidences the adiabatic character of the waveform (\ref{waveform}) and incoherent regime of light-matter interaction. For the kink to preserve self-similarity, the amplitude should be $\Omega_0<1/2T_2$ \cite{Ponomarenko2010}. The numerical calculations of light propagation are based on simulations of the Maxwell-Bloch equations, i.e., Eqs. (\ref{rho})-(\ref{w}) supplemented with the wave equation for the electromagnetic field $\Omega$. The numerical scheme utilized here can be found in the previous publications \cite{Crenshaw1996,Novitsky2009}. Besides those mentioned above, the parameters used in our calculations are $\omega_L=10^{11}$ s$^{-1}$ (so that $\omega_L T_2=0.1$, much less than the threshold for local-field-induced bistability), $n_d=1.5$, $\delta=0$, the light wavelength $\lambda=0.8$ $\mu$m, the medium thickness $L=10 \lambda$.

In order to demonstrate the peculiarities of different regimes of light-matter coupling, we have calculated the dynamics of light transmission and population difference for the input waveform (\ref{waveform}) and different relationships between $t_p$ and relaxation times. Namely, we consider the three regimes as follows:

(i) kink regime, $T_2 \ll t_p \ll T_1$, with $T_2=1$ ps, $T_1=10^3 T_2=1$ ns, $t_p=30 T_2$, and $\Omega_0=0.4/T_2$;

(ii) stationary regime, $T_2, T_1 \ll t_p$, with $T_2=1$ ps, $T_1=10 T_2$, $t_p=30 T_2$, and $\Omega_0=10/T_2$. The larger amplitude $\Omega_0$ is taken, because the low-amplitude radiation as in the case (i) will be mostly absorbed due to its smallness in comparison to the increased saturation amplitude, $\Omega_{sat} \sim (T_1 T_2)^{-1/2}$. This means that we should increase the incoming amplitude to reach closer to saturation again and provide nonzero transmission;

(iii) non-stationary regime, $t_p \sim T_2 \ll T_1$, with $T_2=t_p=1$ ps, $T_1=10^3 T_2$, and $\Omega_0=2.2/T_2$. The increase in the amplitude is again aimed at ensuring approximately the same level of transmission as in the case (i). We use the kink self-similarity condition, which connects the medium thickness with the waveform parameters, $L \sim \Omega_0^2 t_p$ \cite{Novitsky2017}. Since $L$ is unchanged, decrease in $t_p$ should be compensated with increase in $\Omega_0$.

Figure \ref{fig1}(a) shows the transmitted intensity profiles calculated for these three regimes. It is seen that the resulting plateaus of transmission are characterized with approximately the same normalized intensity in all the regimes considered; this fact confirms that these cases, indeed, can be compared. For the kink regime, we see the typical profile with the sharp, shock-wave-like jump of intensity at the front edge due to the self-steepening process. In the stationary regime, on the contrary, the transmitted intensity is smooth, closely following the initial waveform. Finally, in the non-stationary regime, we see the damped oscillations of intensity usual for the abrupt (non-adiabatically, $t_p \lesssim T_1, T_2$) switching external field. These oscillations correspond to the Rabi flopping of the population difference clearly seen in Fig. \ref{fig1}(b) in this case. The period of Rabi oscillations fits well the value of Rabi frequency (although the condition $\Omega_0 T_2 \gg 1$ is not reached producing some discrepancy). There are no such oscillations in the kink and stationary regimes what brings these latter cases together and justifies consideration of the kink regime as a quasi-stationary one and the probable basis for getting quasi-bistable response.

\begin{figure}[t!]
\includegraphics[scale=1., clip=]{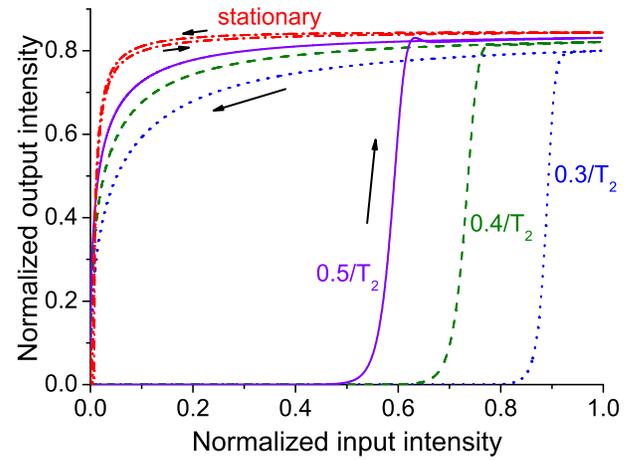}
\caption{\label{fig2} Hysteresis loops calculated from the profiles of Fig. \ref{fig1}(a) for different values of $\Omega_0$ indicated near the corresponding curves.}
\end{figure}

The trailing edge of the transmitted intensity, in contrast to the leading one, has practically the same behavior in all three regimes closely following the incident profile. The difference, or asymmetry, between the dynamics of the leading and trailing edges in the kink regime is the main feature resulting in bistable-like response. Indeed, since the thickness of the medium is small, we can neglect the retardation of waves inside it (in comparison with the kink-like pulse duration) and simply plot the dependence of the transmitted intensity on the incident intensity at the same instant of time. The results of such calculations are shown in Fig. \ref{fig2} (the green dashed curve). For the kink regime, one can see the behavior typical for bistability: there is a jump from low-intensity output to the high-intensity output at a certain value of the growing input (the leading edge), whereas the return to the low-intensity output happens at a completely different, much lower value of the decreasing input (the trailing edge). This difference between the intensity jumps is the direct consequence of the asymmetry between behaviors at the leading and trailing edges of the input waveform. In the stationary case, the curves for the increasing and decreasing inputs almost coincide, so that no bistability occurs [see the red dot-dashed curve in Fig. \ref{fig2}].

Figure \ref{fig2} also shows why we call the observed effect the quasi-bistability and not the true bistability. The true bistability should have non-dynamic origin, i.e., it is not a mere result of optical signal distortion. In particular, for the fixed parameters of the system, the bistable loop should be the same for any input intensity. This is not the case in our kink-based approach: as seen in Fig. \ref{fig2}, changing the waveform amplitude $\Omega_0$ from $0.3/T_2$ to $0.5/T_2$ leads to strong variations of the switching intensity. Nevertheless, the final level of output intensity remains intact depending only on the medium characteristics. This may be used in possible applications, when the specific value of switching intensity is not important and only two pronounced states (on and off) are needed. Keeping this possibility in mind, we call the proposed kink-based mechanism the optical quasi-bistability.

\begin{figure}[t!]
\includegraphics[scale=1., clip=]{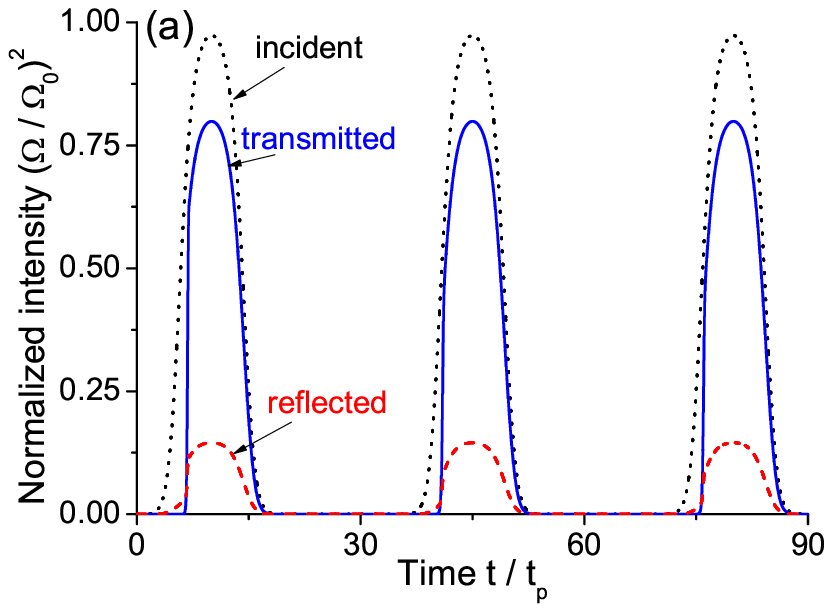}
\includegraphics[scale=1., clip=]{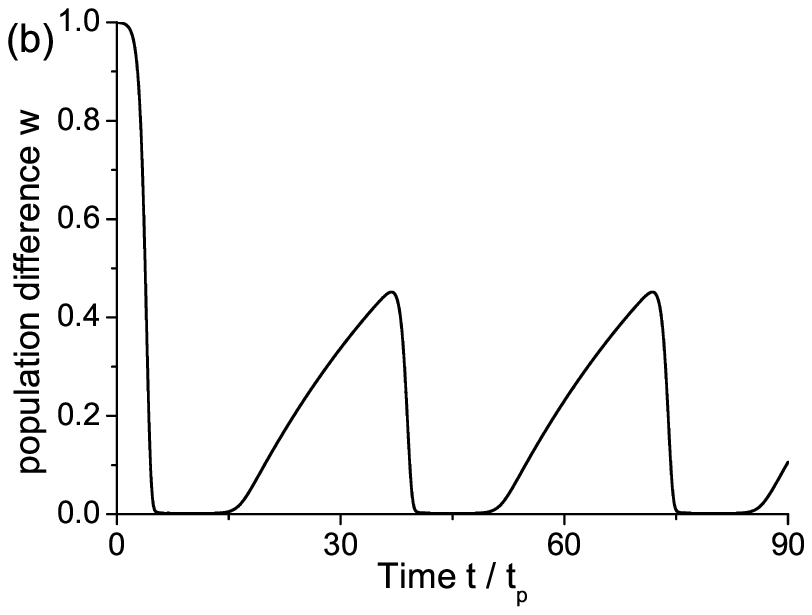}
\caption{\label{fig3} (a) Profiles of transmitted light intensity for the three sequential kinks. (b) Corresponding dynamics of population difference at the entrance of the medium.}
\end{figure}

The time of switching from the low-intensity state to the high-intensity one is governed by $t_p$, which is in our example on the order of several tens of picoseconds. Apart from this switch-on time, it is important to estimate the switch-off time, i.e., the time of switching back from the high-intensity state to the low-intensity one. The switch-off time shows the time interval after which the system can be used again for obtaining the quasi-bistable response. In Fig. \ref{fig3}(a), we demonstrate the transmission and reflection of the three consecutive kink-like pulses with the same parameters as previously, except for the shorter second offset time $t'_0=3 t_0$. We observe the same main features including the characteristic intensity jump at the leading edge of the pulse. These jumps appear for every next waveform launched into the medium after some time interval large enough to preserve the possibility of kink formation. This interval is governed by the relaxation time $T_1$, which determines the return of the medium to the ground state. Figure \ref{fig3}(b) shows that even incomplete return is enough to preserve the consecutive kink formation seen in Fig. \ref{fig3}(a). As a result, we obtain a series of practically identical kinks which can be used for realization of quasi-bistability. In our example, the interval between kinks and, hence, the switch-off time can be estimated as $35 t_p \sim 1$ ns. This corresponds to the frequency of the quasi-bistable operations sequence of the order of $1$ GHz. In fact, the interval between kinks can be further reduced being limited only by the possibility of the next kink formation. One can also consider using other materials, with shorter relaxation times.

\begin{figure}[t!]
\includegraphics[scale=1., clip=]{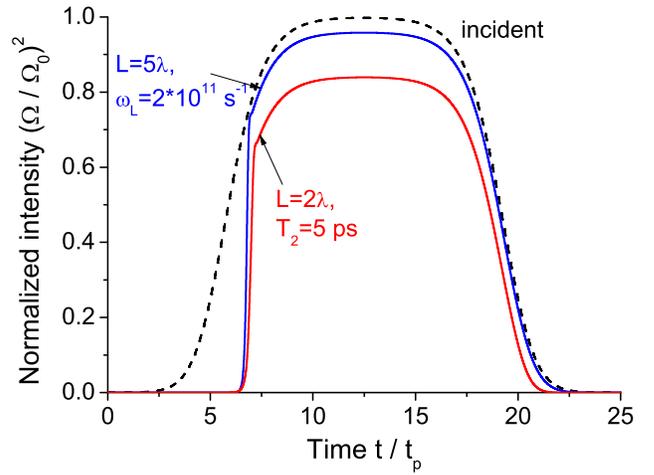}
\caption{\label{fig4} Profiles of transmitted light intensity for the layers of decreased thickness ($5 \lambda$ and $2 \lambda$) and increased Lorentz frequency ($\omega_L=2 \cdot 10^{11}$ s$^{-1}$) and relaxation time ($T_2=5$ ps), respectively.}
\end{figure}

In order to demonstrate the compactness and feasibility of the scheme proposed, we estimate the medium parameters needed. In particular, for quantum dots with large dipole moment $\mu \sim 30$ D \cite{Borri2002} and for the value $\omega_L=10^{11}$ s$^{-1}$ adopted in our calculations, one should take the concentration of $C \sim 3 \cdot 10^{16}$ cm$^{-3}$ which is of the same order as the value used in the self-induced transparency study \cite{Panzarini2002}. Moreover, one can optimize the system parameters using the self-similarity property of the kinks. Indeed, the distance of kink formation is $L= \Omega^2_\infty T^2_2 / \alpha$ \cite{Ponomarenko2010}, where $\Omega_\infty$ is the final kink amplitude and $\alpha$ is the linear absorption coefficient. Since $\Omega_\infty \sim \Omega_0 \sim 1/T_2$ and $\alpha \sim \omega_L T_2$, we have $L \sim 1/\omega_L T_2 \sim 1/\mu^2 C T_2$, i.e., one can make the system even more compact by increasing the density of two-level particles or taking quantum dots with larger dipole moment or relaxation time. As an illustration, Fig. \ref{fig4} shows the profiles of kinks formed after passing the layers of only $5 \lambda$ and even $2 \lambda$ in thickness. The kinks are still present even in these ultracompact instances due to the corresponding tuning of density and relaxation time. This particular behavior opens a way to nanosized all-optical switching devices and logical setups based on quasi-bistable, kink-based processing.

In conclusion, we have found a new mechanism of mirrorless bistable-like response based on the kink formation from incoherent waveform propagating in a dilute resonant medium. We have justified this rather simple mechanism by comparing the kink regime with other regimes of light-matter interaction and demonstrated the dynamic quasi-bistability which can be used for very compact (about two wavelengths in size), low-intensity and fast all-optical switching. This promising effect being free from some drawbacks of existing bistable devices can be successfully utilized for a plenty of highly-demanded applications in optical information processing.

The work was supported by the Belarusian Republican Foundation for Fundamental Research (Project No. F18-049), the Russian Foundation
for Basic Research (Projects No. 18-02-00414, 18-52-00005, and
18-32-00160), the Ministry of Education and Science of
the Russian Federation (GOSZADANIE, Grant No. 3.4982.2017/6.7), and the
Government of Russian Federation (Grant No. 08-08). Numerical
simulations of the nonlinear interaction of light with resonant media have been supported by the Russian Science Foundation (Project No. 18-72-10127).


\begin{thebibliography}{00}
\bibitem{Karim} M.~A.~Karim and A.~A.~Awwal, \textit{Optical Computing: An Introduction} (John Wiley \& Sons, New York, 1992).
\bibitem{Li} X.~Li, Z.~Shao, M.~Zhu, and J.~Yang, \textit{Fundamentals of Optical Computing Technology: Forward the Next Generation Supercomputer} (Springer, Singapore, 2018).
\bibitem{Gibbs} H.~M.~Gibbs, \textit{Optical Bistability: Controlling Light with Light} (Academic Press, Orlando, 1985).
\bibitem{Gibbs1976} H.~M.~Gibbs, S.~L.~McCall, and T.~N.~C.~Venkatesan, {Phys. Rev. Lett.} {\bf36}, 1135 (1976).
\bibitem{Felber1976} F.~S.~Felber and J.~H.~Marburger, {Appl. Phys. Lett.} {\bf28}, 731 (1976).
\bibitem{Winful1979} H.~G.~Winful, J.~H.~Marburger, and E.~Garmire, {Appl. Phys. Lett.} {\bf35}, 379 (1979).
\bibitem{Acklin1993} B.~Acklin, M.~Cada, J.~He, and M.-A.~Dupertuis, {Appl. Phys. Lett.} {\bf63}, 2177 (1993).
\bibitem{Lidorikis2000} E.~Lidorikis and C.~M.~Soukoulis, {Phys. Rev. E} {\bf61}, 5825 (2000).
\bibitem{Soljacic2002} M.~Solja\v{c}i\'{c}, M.~Ibanescu, S.~G.~Johnson, Y.~Fink, and J.~D.~Joannopoulos, {Phys. Rev. E} {\bf66}, 055601(R) (2002).
\bibitem{Teimourpour2012} M.~H.~Teimourpour, {J. Opt.} {\bf14}, 035501 (2012).
\bibitem{Sreekanth2015} K.~V.~Sreekanth, A.~R.~Rashed, A.~Veltri, M.~ElKabbash, and G.~Strangi, {EPL} {\bf112}, 14005 (2015).
\bibitem{Yanik2003} M.~F.~Yanik, Shanhui Fan, and M.~Solja\v{c}i\'{c}, {Appl. Phys. Lett.} {\bf83}, 2739 (2003).
\bibitem{YanikOL2003} M.~F.~Yanik, Shanhui Fan, M.~Solja\v{c}i\'{c}, and J.~D.~Joannopoulos, {Opt. Lett.} {\bf28}, 2506 (2003).
\bibitem{Mingaleev2006} S.~F.~Mingaleev, A.~E.~Miroshnichenko, Yu.~S.~Kivshar, and K.~Busch, {Phys. Rev. E} {\bf74}, 046603 (2006).
\bibitem{Kabakova2010} I.~V.~Kabakova, C.~M.~de~Sterke, and B.~J.~Eggleton, {J. Opt. Soc. Am. B} {\bf27}, 2648 (2010).
\bibitem{Zhang2010} W.~L.~Zhang and S.~F.~Yu, {Opt. Commun.} {\bf283}, 2622 (2010).
\bibitem{Lapine2014} M.~Lapine, I.~V.~Shadrivov, and Yu.~S.~Kivshar, {Rev. Mod. Phys.} {\bf86}, 1093 (2014).
\bibitem{Tuz2010} V.~R.~Tuz, S.~L.~Prosvirnin, and L.~A.~Kochetova, {Phys. Rev. B} {\bf82}, 233402 (2010).
\bibitem{Tuz2012} V.~R.~Tuz, V.~S.~Butylkin, and S.~L.~Prosvirnin, {J. Opt.} {\bf14}, 045102 (2012).
\bibitem{Kim2018} M.~Kim, S.~Kim, and S.~Kim, {Opt. Express} {\bf26}, 11620 (2018).
\bibitem{Shadrivov2010} I.~V.~Shadrivov, K.~Y.~Bliokh, Yu.~P.~Bliokh, V.~Freilikher, and Yu.~S.~Kivshar, {Phys. Rev. Lett.} {\bf104}, 123902 (2010).
\bibitem{Yuan2016} H.~Yuan, X.~Jiang, F.~Huang, and X.~Sun, {Opt. Lett.} {\bf41}, 661 (2016).
\bibitem{Soljacic2003} M.~Solja\v{c}i\'{c}, M.~Ibanescu, C.~Luo, S.~G.~Johnson, S.~Fan, Y.~Fink, and J.~D.~Joannopoulos, {Proc. SPIE} {\bf5000}, 200 (2003).
\bibitem{Rozanov1981} N.~N.~Rozanov, {J. Exp. Theor. Phys.} {\bf53}, 47 (1981).
\bibitem{Miller1984} D.~A.~B.~Miller, A.~C.~Gossard, and W.~Wiegmann, {Opt. Lett.} {\bf9}, 162 (1984).
\bibitem{Gribkovskii1988} V.~P.~Gribkovskii, L.~G.~Zimin, S.~V.~Gaponenko, I.~E.~Malinovskii, P.~I.~Kuznetsov, and G.~G.~Yakushcheva, {phys. stat. sol. b} {\bf150}, 761 (1988).
\bibitem{Hopf1984} F.~A.~Hopf, C.~M.~Bowden, and W.~H.~Louisell, {Phys. Rev. A} {\bf29}, 2591 (1984).
\bibitem{Ben-Aryeh1986} Y.~Ben-Aryeh, C.~M.~Bowden, and J.~C.~Englund, {Phys. Rev. A} {\bf34}, 3917 (1986).
\bibitem{Haus1988} J.~W.~Haus, L.~Wang, M.~Scalora, and C.~M.~Bowden, {Phys. Rev. A} {\bf38}, 4043 (1988).
\bibitem{Friedberg1989} R.~Friedberg, S.~R.~Hartmann, and J.~T.~Manassah, {Phys. Rev. A} {\bf39}, 3444 (1989).
\bibitem{Benedict1991} M.~G.~Benedict, V.~A.~Malyshev, E.~D.~Trifonov, and A.~I.~Zaitsev, {Phys. Rev. A} {\bf43}, 3845 (1991).
\bibitem{Malyshev1997} V.~Malyshev and E.~C.~Jarque, {J. Opt. Soc. Am. B} {\bf14}, 1167 (1997).
\bibitem{Afanas'ev1998} A.~A.~Afanas'ev, R.~A.~Vlasov, N.~B.~Gubar, and V.~M.~Volkov, {J. Opt. Soc. Am. B} {\bf15}, 1160 (1998).
\bibitem{Afanas'ev1999} A.~A.~Afanas'ev, A.~G.~Cherstvy, R.~A.~Vlasov, and V.~M.~Volkov, {Phys. Rev. A} {\bf60}, 1523 (1999).
\bibitem{Brunel1999} M.~Brunel, C.~\"{O}zkul, and F.~Sanchez, {J. Opt. Soc. Am. B} {\bf16}, 1886 (1999).
\bibitem{Novitsky2008} D.~V.~Novitsky and S.~Yu.~Mikhnevich, {J. Opt. Soc. Am. B} {\bf25}, 1362 (2008).
\bibitem{Hehlen1994} M.~P.~Hehlen, H.~U.~G\"{u}del, Q.~Shu, J.~Rai, S.~Rai, and S.~C.~Rand, {Phys. Rev. Lett.} {\bf73}, 1103 (1994).
\bibitem{Paspalakis2006} E.~Paspalakis, A.~Kalini, and A.~F.~Terzis, {Phys. Rev. B} {\bf73}, 073305 (2006).
\bibitem{Malyshev1996} V.~Malyshev and P.~Moreno, {Phys. Rev. A} {\bf53}, 416 (1996).
\bibitem{Gibbs1985} H.~M.~Gibbs, G.~R.~Olbright, N.~Peyghambarian, H.~E.~Schmidt, S.~W.~Koch, and H.~Haug, {Phys. Rev. A} {\bf32}, 692 (1985).
\bibitem{Lindberg1986} M.~Lindberg, S.~W.~Koch, and H.~Haug, {Phys. Rev. A} {\bf33}, 407 (1986).
\bibitem{Ponomarenko2010} S.~A.~Ponomarenko and S.~Haghgoo, {Phys. Rev. A} {\bf82}, 051801(R) (2010).
\bibitem{Novitsky2017} D.~V.~Novitsky, {Phys. Rev. A} {\bf95}, 053846 (2017).
\bibitem{Bowden1993} C.~M.~Bowden and J.~P.~Dowling, {Phys. Rev. A} {\bf47}, 1247 (1993).
\bibitem{Crenshaw2008} M.~E.~Crenshaw, {Phys. Rev. A} {\bf78}, 053827 (2008).
\bibitem{Crenshaw1996} M.~E.~Crenshaw, {Phys. Rev. A} {\bf54}, 3559 (1996).
\bibitem{Novitsky2009} D.~V.~Novitsky, {Phys. Rev. A} {\bf79}, 023828 (2009).
\bibitem{Borri2002} P.~Borri, W.~Langbein, S.~Schneider, U.~Woggon, R.~L.~Sellin, D.~Ouyang, and D.~Bimberg, {Phys. Rev. B} {\bf66}, 081306(R) (2002).
\bibitem{Panzarini2002} G.~Panzarini, U.~Hohenester, and E.~Molinari, {Phys. Rev. B} {\bf65}, 165322 (2002).
\end{thebibliography}
\end{document}